\documentstyle[twoside,fleqn,espcrc2,epsf,epsfig]{article}
\newcommand{\AmS}{{\protect\the\textfont2
  A\kern-.1667em\lower.5ex\hbox{M}\kern-.125emS}}

\newcommand{\nn}{\nonumber}
\newcommand{\be}{\begin{equation}}
\newcommand{\ee}{\end{equation}}
\newcommand{\bea}{\begin{eqnarray}}
\newcommand{\eea}{\end{eqnarray}}

\def\als{\alpha_{s}}
\def\siml{{\
    \lower-1.2pt\vbox{\hbox{\rlap{$<$}\lower6pt\vbox{\hbox{$\sim$}}}}\ }}

\title{Phenomenological impact of the 
resummation of logs of $\alpha$ in heavy quarkonium} 

\author{Antonio Pineda\address{Dept. d'Estructura i Constituents de la
  Mat\`eria, U. Barcelona, Diagonal 647, E-08028 Barcelona, Spain} 
  \thanks{This work was supported in part by
MCyT and Feder, FPA2001-3598, by CIRIT, 2001SGR-00065
and by the EU network EURIDICE, HPRN-CT2002-00311.}}

\begin{document}

\begin{abstract}
Here we would like to review recent progress on the resummation 
on logarithms of $\als$ in heavy quarkonium. 
We will mainly focus on the phenomenological 
relevance of these achievements. Determinations of the $\eta_b(1S)$ mass, 
$B_c(1S)$ hyperfine splitting, inclusive electromagnetic decays and 
implications for $t$-$\bar t$ production near threshold. 
\end{abstract}

\maketitle

\section{Introduction}

\noindent
Heavy quark-antiquark systems near threshold are characterized by the
small relative velocity $v$ of the heavy quarks in their center of
mass frame. This small parameter produces a hierarchy of widely
separated scales: $m$ (hard), $mv$ (soft), $mv^2$ (ultrasoft), ... 
The factorization between them is efficiently achieved by using
effective field theories, where one can organize the calculation as
various perturbative expansions on the ratio of the different scales
effectively producing an expansion in $v$. The terms in these series
get multiplied by parametrically large logs: $\ln v$, which can also
be understood as the ratio of the different scales appearing in the
physical system. Again, effective field theories are very efficient in
the resummation of these large logs once a non-relativistic 
renormalization group (NRG)
analysis of them has been performed. We will review in this paper 
recent progress on the resummation of the above logarithms, 
within the context of pNRQCD \cite{pNRQCD}, in the weak coupling regime 
($v \sim \als$). Besides the pure theoretical 
interest of these computations, they may also have an important 
phenomenological impact in several situations. Let us enumerate a few of 
them. The determination of the bottom and charm masses (using the 
experimental value of the ground state heavy quarkonium masses or 
non-relativistis sum rules). The 
determination of the $\eta_b(1S)$ mass, the hyperfine splitting (HFS) of 
the ground state $B_c$ system, or theoretical improved determinations of  
the $\eta_c$. One can also try to obtain improved determinations 
for the inclusive electromagnetic decays of the heavy quarkonium. 
On the other hand the application of this program to 
$t$-$\bar t$ production near threshold at the Next Linear Collider ´
is one of the main 
motivations to undergo these computations.  
In this paper we review the phenomenological analysis already available 
in the literature and outline possible future work.

\section{Hyperfine splitting}

Analytical expressions for the HFS of heavy quarkonium 
(for the equal and non-equal mass case)
are available with leading log (LL) \cite{HMS,RGmass} and 
next-to-leading log (NLL) \cite{KPPSS,Penin:2004xi} 
accuracy. 

\begin{figure}[t]
\epsfxsize=3in
\epsffile{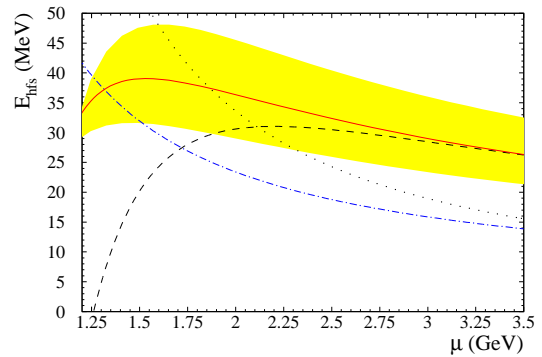}
\caption{\label{fig1}{\it  HFS of 1S bottomonium as a function of the
renormalization scale $\mu$ in the LO (dotted line), NLO (dashed line), LL
(dot-dashed line), and NLL (solid line) approximations. For the NLL
result, the band reflects the errors due to $\alpha_s(M_Z)=0.118\pm
0.003$.}}
\end{figure}

For the case of bottomonium, these results have 
been used in Ref. \cite{KPPSS} to give predictions for the mass 
of the $\eta_b(1S)$ (using the 
experimentally very well known value of $M_{\Upsilon(1S)}$) with 
great precision
\begin{equation}
M(\eta_b(1S))=9421\pm 11\,{(\rm th)} \,{}^{+9}_{-8}\, 
(\delta\alpha_s)~{\rm MeV}\,,
\end{equation}
where the errors due to the high-order perturbative corrections and the
nonperturbative effects are added up in quadrature in ``th'', whereas
``$\delta\alpha_s$'' stands for the uncertainty in
$\alpha_s(M_Z)=0.118\pm0.003$. This prediction is of great experimental 
interest. The discovery of
the $\eta_b$ meson is one of the primary goals of the CLEO-c research program
\cite{Sto} and there are experimental proposals for its detection 
at Tevatron too \cite{Maltoni:2004hv}. It has also been argued that the HFS 
can be used to search for new physics \cite{Sanchis-Lozano:2004gh}. 
Therefore, an accurate prediction of its 
mass $M(\eta_b)$ is thus a big challenge and a test for the QCD theory of 
heavy quarkonium. For instance, this prediction 
can be compared with those obtained
either in lattice \cite{Eic},  potential models (see for instance 
Ref.~\cite{GodRos}) or 
sum rules \cite{Narison}. It seems to be a
general trend that our result is larger than the lattice predictions and
smaller than most of the potential model results.
We would also like to remark that the inclusion of resummation of 
logarithms has a sizable effect in the determination of the $\eta_b(1S)$
mass. We illustrate this fact in Fig. \ref{fig1} from Ref. \cite{KPPSS}. 
In this figure, the HFS for the bottomonium
ground state is plotted as a function of $\mu$ in the LO, NLO, LL, and NLL
approximations. As we see, the LL curve shows a weaker scale dependence
compared to the LO one. The scale dependence of the NLO and NLL expressions is
further reduced, and, moreover, the NLL approximation remains stable up to
smaller scales than the fixed-order calculation. At the scale $\mu'\approx
1.3$~GeV, which is close to the inverse Bohr radius, the NLL correction
vanishes.  Furthermore, at $\mu''\approx 1.5$~GeV, where
$\alpha_s^{LL}=0.319$,  the result becomes
independent of $\mu$; {\it i.e.}, the NLL curve shows a local maximum.  
This suggests a nice convergence of the logarithmic expansion despite the presence
of the ultrasoft contribution with $\alpha_s$ normalized at the rather low
scale $\bar\mu^2/m_b\sim 0.8$~GeV.  By taking the difference of the NLL and LL
results at the local maxima as a conservative estimate of the error due to
uncalculated perturbative higher-order contributions, one gets $E_{\rm hfs}=39\pm 8$~MeV. A
similar error estimate is obtained by the variation of the 
normalization scale in the physically motivated soft region $1-3$ GeV.

\begin{figure}[t]
\begin{center}
\epsfxsize=3in 
\epsffile{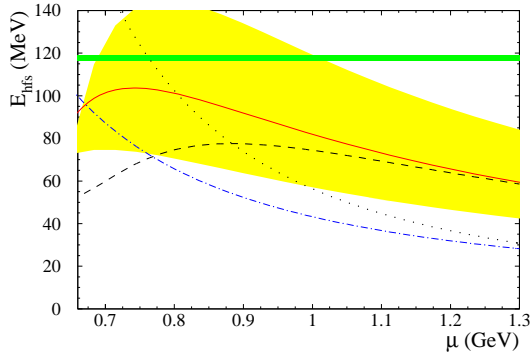}
\end{center}
\caption{\label{fig2} {\it HFS of 1S charmonium as a function of the
renormalization scale $\mu$ in the LO (dotted line), NLO (dashed line), LL
(dot-dashed line), and NLL (solid line) approximations. 
For the NLL result, the band reflects the errors
due to $\alpha_s(M_Z)=0.118\pm 0.003$. The horizontal band gives 
the experimental value $117.7\pm 1.3$~MeV \cite{Hag}.}}
\end{figure}

For the case of charmonium, the use of perturbation theory is more doubtful, 
even though there has been some attempts in this direction recently 
\cite{BSV} (see also \cite{BraVai} for the $B_c$), 
yet one can try and see what comes out. This is specially important since 
even unquenched attempts to obtain the HFS of charmonium from 
lattice undershot the experimental value by around $20\%$ \cite{Pierro}. 
The results obtained in Ref. \cite{KPPSS} are given in
Fig.~\ref{fig2} along with the experimental value $117.7\pm 1.3$~MeV
\cite{Hag}. The local maximum of the NLL curve corresponds to $E_{\rm
hfs}=104$~MeV and  $\alpha_s^{LL}=0.534$.  We should emphasize the crucial role of the resummation to
bring the perturbative prediction closer to the experimental
figure. Therefore, in Ref. \cite{KPPSS} the whole difference 
of $\approx 14$ MeV between the perturbative 
prediction and the experimental value for the HFS of the 
ground state of charmonium was used to estimate the size of the non-perturbative 
effects. In any case,
within the power counting assumed in Ref. \cite{KPPSS}, these non-perturbative
effects are beyond the accuracy of this computation and are added to the
errors. Taking into account that they
are suppressed by the inverse heavy-quark mass as $1/(\alpha_sm_q)^2$ due 
to the multipole expansion, 
one obtains $\approx 3.5$ MeV for the typical size of the
nonperturbative contribution to the HFS in bottomonium. For the estimate of
the nonperturbative error, this number was multiplied by two, which 
was used above for the determination of the theoretical error. 
These formulae has also been applied to $n=2$ excited states. 
For bottomonium, one obtains $E_{\rm
hfs}(2S)/E_{\rm hfs}(1S)=0.25$.  For charmonium, our perturbative
estimate $E_{\rm hfs}(2S)/E_{\rm hfs}(1S)=0.37$ also reasonably agrees with
the result $0.41\pm 0.03$ of the recent experimental measurements
\cite{Cho}. Although one cannot rely on the (even NRG-improved)
perturbative analysis of the excited charmonium states, the above agreement
suggests that the nonperturbative effects are small, at least for the ground
state.

\begin{figure}[t]
\begin{center}
\epsfxsize=3in
\epsffile{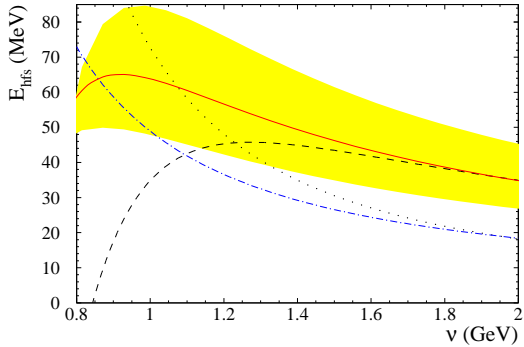}
\end{center}
\caption{\label{fig3}´{\it HFS for charm-bottom quarkonium as the function of the
renormalization scale $\nu$ in LO (dotted line), NLO (dashed line), LL
(dot-dashed line), and NLL (solid line) approximation for $\nu_h=1.95$~GeV.
For the NLL result the band reflects the errors due to $\alpha_s(M_Z)=0.118\pm
0.003$.}}
\end{figure}

It  has also been possible to give a good prediction for the HFS 
of the $B_c(1S)$ system\footnote{In this case it is the 
pseudoscalar, $B_c(1S)$, which 
has been measured, nevertheless not with enough precision to allow a good 
absolute determination of the $B_c^*(1S)$ mass. However, 
with improved statistics and precision of the
$B_c$ data, this result can be considered as a prediction for the $B_c^*$ meson
mass.} \cite{Penin:2004xi}
\be
M(B^*_c)-M(B_c)=65 \pm
24\,{(\rm th)}\,{}^{+19}_{-16}\,(\delta\alpha_s)~{\rm MeV}
\ee
Potential models appear to obtain slightly larger numbers \cite{Bcpotential}.
As in the previous cases the inclusion of logarithms appears to be a
large effect. 
In
Fig.~\ref{fig3} from ref. \cite{Penin:2004xi}, the HFS for the charm-bottom 
quarkonium ground state is
plotted as a function of $\nu$ in the LO, NLO, LL, and NLL approximations for
the hard matching scale value $\nu_h= 1.95$~GeV.  As we see, the LL curve
shows a weaker scale dependence compared to the LO one.  The scale dependence
of the NLO and NLL expressions is further reduced, and, moreover, the NLL
approximation remains stable at the physically motivated scale of the inverse
Bohr radius, $C_F\alpha_s m_r\sim 0.9$~GeV, where the fixed-order expansion
breaks down. At the scale $\nu'\approx 0.85$~GeV, which is close to the inverse
Bohr radius, the NLL correction vanishes.  Furthermore, at $\nu''=
0.92$~GeV, the result becomes independent of $\nu$; {\it i.e.}, the NLL curve
shows a local maximum corresponding to $E_{\rm hfs}=65$~MeV, which is taken as
the central value of the estimate.  The NLL curve also shows an impressive
stability with respect to the hard matching scale variation in the physical
range $m_c<\nu_h<m_b$ and has a
local maximum at $\nu_h= 1.95$~GeV, which is taken for the numerical estimates.
All this suggests a nice convergence of the logarithmic expansion despite the
presence of the ultrasoft contribution where $\alpha_s$ is normalized at the
rather low scale $\bar\nu^2/\nu_h\sim 0.5$~GeV.
Let us discuss the accuracy of our result.  For a first estimate of the error
due to uncalculated higher-order contributions, we take $9$~MeV, the
difference of the NLL and LL results at the local maxima.  A different
estimate can be obtained by varying the normalization scale in the physical
range $0.8\le\nu\le 1.4$~GeV. In this case the difference with the maximum is
$16$~MeV. Being conservative, we take this second number for our estimate of
the perturbative error. Within the power counting assumed in Ref. 
\cite{Penin:2004xi}, the
nonperturbative effects are beyond the accuracy of the computation and 
added to the errors. They are also inferred using charmonium
data in the same way than above and $\approx 18$
MeV for the error due to the nonperturbative contribution to the HFS in
$B_c$ was obtained.

\section{Production and Annihilation rates}

There has also been much progress on the resummation of large logarithms 
appearing in the 
annihilation of a heavy quarkonium into leptons, photons or light
hadrons, as well as its production in $e^+e^-$ or $\gamma\gamma$
collisions. 

\begin{figure}[t]
\begin{center}
\epsfxsize=3in
\epsffile{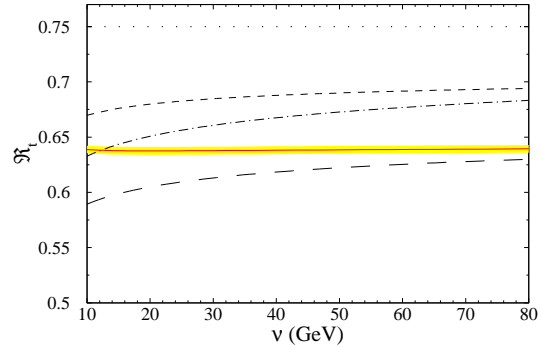}
\end{center}
\caption{\label{figt} {\it The spin  ratio  as the function of the
renormalization scale $\nu$ in LO$\equiv$LL (dotted line), NLO (short-dashed
line), NNLO (long-dashed line), NLL (dot-dashed
line), and NNLL (solid line) approximation for the (would be)
toponium ground state with $\nu_h=m_t$. For the NNLL result the band
reflects the errors due to $\alpha_s(M_Z)=0.118\pm 0.003$. Note that
for the vertical axis the zero is suppressed.}}
\end{figure}

\begin{figure}[t]
\begin{center}
\epsfxsize=3in
\epsffile{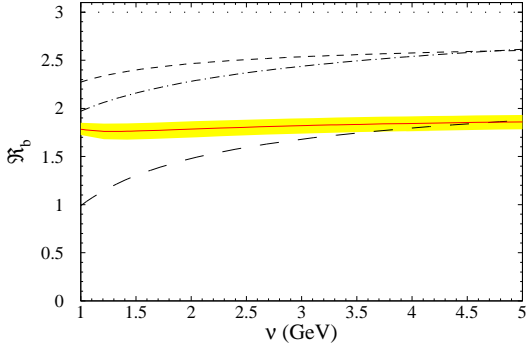}
\end{center}
\caption{\label{figb} {\it The spin  ratio  as the function of the
renormalization scale $\nu$ in LO$\equiv$LL (dotted line), NLO (short-dashed
line), NNLO (long-dashed line), NLL (dot-dashed
line), and NNLL (solid line) approximation for the bottomonium
ground state with $\nu_h=m_b$. For the NNLL result the band reflects
the errors due to $\alpha_s(M_Z)=0.118\pm 0.003$.}}
\end{figure}

\begin{figure}[t]
\begin{center}
\epsfxsize=3in
\epsffile{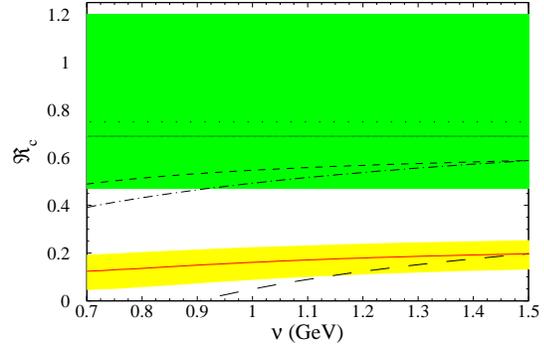}
\end{center}
\caption{\label{figc} {\it The spin  ratio  as the function of the
renormalization scale $\nu$ in LO$\equiv$LL (dotted line), NLO (short-dashed
line), NNLO (long-dashed line), NLL (dot-dashed
line), and NNLL (solid line) approximation for the charmonium
ground state with $\nu_h=m_c$.  For the NNLL result the lower (yellow)
band reflects
the errors due to $\alpha_s(M_Z)=0.118\pm 0.003$. The upper (green) band
represents the experimental error of the ratio \cite{Hag} where the central 
value is given by the horizontal solid line.}}
\end{figure}

The 
resummation of  the large logarithms of the heavy quark velocity to
all orders in $\alpha_s$  has been advocated as a tool to improve the behaviour
of the perturbative expansion for $t\bar t$ threshold production \cite{HMST}.
Currently, the complete next-to-leading logarithmic (NLL) approximation
for the production and annihilation rates is available
\cite{csNLL,Hoang:2002yy}. In Ref. \cite{Pin3} a phenomenological 
analysis of the NLL result for the electromagnetic inclusive decays of the 
heavy quarkonium was made. The first attempt to go beyond the NLL
approximation \cite{HMST} suggested a very good convergence of the
logarithmic expansion.  In particular, an accuracy of 2-3\% was 
claimed for the cross section of $t\bar t$ threshold production.
However, subsequent calculations of some   
next-to-next-to-leading logarithmic (NNLL)
terms \cite{Hoa}, which had not been taken into account in
Ref.~\cite{HMST}, 
casted serious doubts on this estimate. Thus, the full
calculation of the NNLL corrections, which still remains elusive, 
is unavoidable to draw definite conclusions. In Ref. \cite{Penin:2004ay},
it has recently been derived the {\it complete} NNLL result for the
spin dependent part of the heavy quarkonium production annihilation
rates, which includes the terms of the form $\alpha_s^{n+2}\ln^n\alpha_s$
for all $n$ and applied to the heavy quarkonium phenomenology. 
In Figs.~\ref{figt}, \ref{figb} and \ref{figc}, the spin ratio
is plotted as a function of $\nu$ in the various logarithmic and 
fixed-order approximations for the (hypothetical) toponium, bottomonium and
charmonium ground states, respectively.  As we see, in the second order the
convergence and stability of the result with respect to the scale
variation is substantially improved if one switches from the
fixed-order to the logarithmic expansion. We want to remark that the
$\nu$ dependence of the NLL approximation is slightly worse than at
NLO. This is due to the artificially small $\nu$ dependence at NLO
which is likely due to the fact that at this order only the hard scale
enters.

Let us first consider the top quark case. In particular,
 the ratio of the cross sections of the
resonance $e^+e^-\to t\bar t$ and $\gamma\gamma\to t\bar t$ production.
As one can see in Fig.~\ref{figt}, 
the logarithmic expansion shows perfect convergence and the NNLL
correction vanishes at the scale $\nu\approx 13$~GeV, which is close to
the physically motivated scale of the inverse Bohr radius
$\alpha_sm_t/2$. For illustration, at the scale of minimal sensitivity,
$\nu=20.2$, GeV we have
\be
{\sigma_{\rm res}(e^+e^-\rightarrow  t\bar t)\over\sigma_{\rm res}
(\gamma\gamma\to t\bar t)}={1\over 3Q_t^2}\left(1-0.132-0.018\right)
\,.
\ee 
However, it is not clear if the nice behaviour of the
logarithmic expansion also holds for the spin-independent part of the
threshold cross section. A possible problem is connected to 
the ultrasoft contribution,
which is enhanced by the larger value of $\alpha_s$ at the ultrasoft
scale. Whereas it is suppressed in the spin ratio by the fifth power of $\alpha_s$, 
for the spin-independent part it already contributes at  
${\cal O}(\alpha_s^3)$ and can destabilize the expansion.

For bottomonium, the logarithmic expansion shows nice convergence and
stability (c.f. Fig.~\ref{figb})
despite the presence of ultrasoft contributions with
$\alpha_s$ normalized at a rather low scale $\nu^2/m_b$. 
At the same time, the perturbative corrections are important
and reduce the leading order result by approximately $41\%$.  
For illustration, at the
scale of minimal sensitivity, $\nu=1.295$~GeV, we have the following series:
\be
{\Gamma(\Upsilon(1S) \rightarrow
  e^+e^-)\over\Gamma(\eta_b(1S)\rightarrow\gamma\gamma)} 
={1\over 3Q_b^2}\left(1-0.302-0.111\right)\,.
\ee 
In contrast, the fixed-order expansion blows up at the scale
of the inverse Bohr radius.

\begin{figure}[h]
\hspace{-0.1in}
\epsfxsize=2.8in
\centerline{
\epsffile{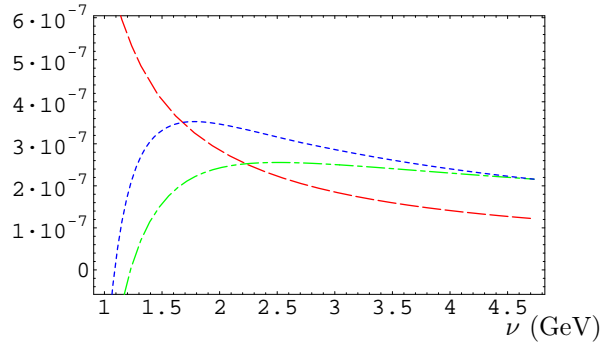}
\put(-15,-3){$\nu$ (GeV)}
}
\caption {{\it Plot of $\Gamma(\eta_b (1S) \rightarrow \gamma\gamma)$
with LO/LL (dashed line), NLO (dot-dashed line) and NLL (dotted line)
accuracy versus the renormalization scale $\nu$.}}
\label{etab1SggNLL}
\end{figure}

So far we have discussed the perturbative corrections to the Coulomb-like
quarkonium.  However, in contrast to the $t\bar{t}$ system, 
for bottomonium nonperturbative contributions can be
important. In our case the interaction of the quark-antiquark pair to the
nonperturbative gluonic field is suppressed by $v$ through
the multipole expansion in the same way than for the HFS computation.
In this case however they only contribute in the N$^4$LL approximation, 
far beyond the precision of this computation. Note that the nonperturbative contribution to the
decay rates ratio
is suppressed by a factor $v^2$ in comparison to the
binding energy and decay rates, where the leading nonperturbative effect
is due to chromoelectric dipole interaction.  Thus, by using the 
available experimental data on the
$\Upsilon$ meson as input, one can predict the production and
annihilation rates of the yet undiscovered $\eta_b$ meson. In particular, 
one can predict the $\eta_b(1S)$ decay rate using
the experimental value for the $\Upsilon(1S)$ decay rate and the following 
figure was obtained in Ref. \cite{Penin:2004ay}
\bea
\nn
\label{etabgg}
\Gamma(\eta_b(1S) \rightarrow \gamma\gamma)&=&0.659\pm 0.089 ({\rm th.}) 
{}^{+0.019}_{-0.018} (\alpha_{\rm s})
\\
&&
\pm 0.015 ({\rm exp.})\; {\rm keV}
\,,
\eea
where $\nu=1.295$ GeV, the scale of minimal sensitivity, was taken 
for the central value, the difference between the NLL and NNLL result
for the theoretical error and $\alpha_{\rm s}(M_Z)=0.118 \pm 0.003$. The
last error in Eq.~(\ref{etabgg}) reflects the experimental error of
$\Gamma(\Upsilon(1S) \rightarrow e^+e^-)=1.314\pm 0.029$ keV \cite{Hag}.
This value considerably exceeds the result for the absolute value of the
decay width obtained in Ref.~\cite{Pin3} on the basis of the full NLL
analysis including the spin-independent part (see Fig. \ref{etab1SggNLL}):
\be
\Gamma(\eta_b (1S) \rightarrow \gamma\gamma)=0.35 \pm 0.1 ({\rm th.})
\pm 0.05 (\alpha_{\rm s}) {\rm KeV}.
\ee 
This can be a signal of
slow convergence of the logarithmic expansion for the spin-independent
contribution which is more sensitive to the dynamics of the bound state 
and in particular to the ultrasoft contribution as it has been 
discussed above. On the other hand, the renormalon effects  \cite{BC}
could produce some systematic errors in the pure perturbative
evaluations of the production/annihilation rates. The problem is
expected to be more severe for the charmonium case discussed below.

We would also like to remark that the one-loop result for $\nu=m_b$
overshoots the NNLL one by approximately $30\%$.  This casts some doubts
on the accuracy of the existing $\alpha_s$ determination from the
$\Gamma(\Upsilon\to{\rm light~hadrons})/\Gamma(\Upsilon\to e^+e^-)$
decay rates ratio, which gives $\alpha_s(m_b)=0.177\pm 0.01$, well below
the ``world average'' value \cite{Hag}.  The theoretical uncertainty in
the analysis is estimated through the scale dependence of the one-loop
result.  Our analysis of the photon mediated annihilation rates
indicates that the actual magnitude of the higher order corrections is
most likely quite beyond such an estimate and the theoretical
uncertainty given in \cite{Hag} should be increased by a factor of
two. This brings the result for $\alpha_s$ into $1\sigma$ distance from
the ``world average'' value.

\begin{figure}[h]
\hspace{-0.1in}
\epsfxsize=2.8in
\centerline{
\epsffile{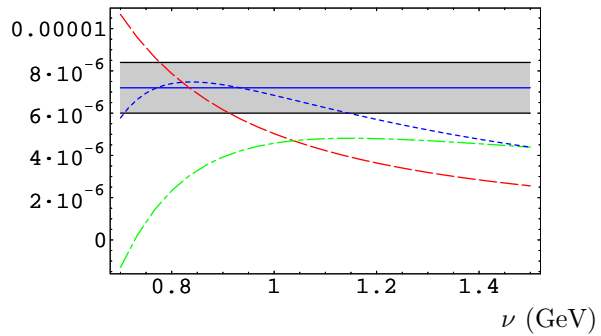}
\put(-15,-3){$\nu$ (GeV)}
}
\caption {{\it Plot of $\Gamma(\eta_c (1S) \rightarrow \gamma\gamma)$ 
with LO/LL (dashed line), NLO (dot-dashed line) and NLL (solid line) accuracy   
versus the renormalization scale $\nu$. The horizontal line and its
band give the experimental value and its errors: $\Gamma(\eta_c (1S) \rightarrow \gamma\gamma)=7.2 \pm 1.2$ KeV \cite{Hag}.}}
\label{etac1Sgg}
\end{figure}

For the charmonium, the NNLO approximation becomes negative at an
intermediate scale between $\alpha_sm_c$ and $m_c$ (c.f. Fig.~\ref{figc})
and the use of the
NRG is mandatory to get a sensible perturbative approximation. The NNLL
approximation has good stability against the scale variation but the
logarithmic expansion does not converge well.  This is the main factor
that limits the theoretical accuracy since the nonperturbative
contribution is expected to be under control. For illustration, at the
scale of minimal sensitivity, $\nu=0.645$ GeV, one obtains \cite{Penin:2004ay}
\be
{\Gamma(J/\Psi(1S)\rightarrow
  e^+e^-)\over\Gamma(\eta_c(1S)\rightarrow\gamma\gamma)} 
={1\over 3Q_c^2}\left(1-0.513-0.326\right)\,.
\ee 
The central value is $2\sigma$ below the experimental
one. The discrepancy may be explained by the large higher order
contributions. This should not be surprising because of the rather large value
of $\alpha_s$ at the inverse Bohr radius of charmonium.  For the charmonium
HFS, however, the logarithmic expansion converges well and the
prediction of the NRG is in perfect agreement with the experimental data.
Thus one can try to improve the convergence of the series for the
production/annihilation rates by accurately taking into account the
renormalon-related contributions. One point to note is that with a potential
model evaluation of the wave function correction, the sign of the NNLO term is
reversed in the charmonium case \cite{CzaMel2}. At the same time the
subtraction of the pole mass renormalon from the perturbative static potential
makes explicit that the potential is steeper and closer to lattice and
phenomenological potential models \cite{sum}. Therefore, the incorporation of
higher order effects from the static potential may improve the agreement with
experiment. Finally, 
we can not avoid mention that the NLL evaluation for the decay is able 
to reproduce the experimental value (see Fig. \ref{etac1Sgg}). 
Therefore, some extra work needs to be done to clarify these issues.

\section{Conclusions and outlook} 

We have seen that the resummation of logarithms appear to have a large 
phenomenological impact in the heavy quarkonium physics. This is so in 
top-, bottom- or charmonium physics. 

First, the heavy-quarkonium HFS have been studied in the NLL
approximation. The use of the NRG extends the range of
$\mu$ where the perturbative result is stable to the physical scale of the
inverse Bohr radius. The resummation of logarithms is found to be crucial to
bring the perturbative prediction closer to the experimental figure of the HFS
in charmonium (despite {\it a priori} unsuppressed nonperturbative effects), 
and to give reliable predictions for the $\eta_b(1S)$ mass and the 
$B_c(1S)$ HFS. 
These results seem to indicate that the properties of the charmonium, $B_c$ and
bottomonium ground states are dictated by perturbative dynamics.  

In the case of $t$-$\bar t$ 
production near threshold, the partial NNLL analysis
made in Ref. \cite{Hoa} does not seem to show a very nice convergence 
(even if the absolute value of the corrections is small). Nevertheless,
being incomplete, such analysis is scheme dependent. 
In Ref. 
\cite{Penin:2004ay} a complete result with NNLL accuracy has been obtained for the
ratio of the spin one and spin zero production. This is a physical result 
by itself and therefore scheme independent. In this case a very nice convergence is 
found. Nevertheless, in this case the contribution due to the ultrasoft 
scale is suppressed. Therefore, it is premature to draw any definite 
conclusion for the convergence of the series.
We are then eagerly waiting for the complete 
NNLL evaluation, which, even if difficult, is within reach. 
This is of utmost importance
for the future determinations of the top mass and the Higgs-top 
coupling at the Next Linear Collider \cite{MarMiq}. 

For the inclusive electromagnetic decays of the bottomonium and 
charmonium ground states the effects due to the resummation of logarithms 
appear to be large and always improve the result compared with the finite 
order evaluations, yet the errors are large and further work seems to be 
needed. Here as well, the complete NNLL evaluation would be of utmost 
importance to further clarify the physical picture. 
However, one should not forget the possible drawbacks of these determinations. 
There is an implicit dependence on the ultrasoft scale, which for the 
case of bottomonium and charmonium is quite low and is related with 
non-perturbative effects, which should be eventually studied with the 
help of lattice or models. Moreover, renormalon effects could 
also play a role. 

Further work is ahead. No analysis has been made yet for the 
non-relativistic sum rules (to obtain them with NNLL accuracy 
would be a major step to obtain accurate determinations of the bottom 
and charm masses since they are strongly scale dependent), 
nor the impact of the resummation of 
logarithms in the determination of the bottom and charm masses from the 
ground state masses (which are known with NNLL accuracy 
\cite{RGmass,Hoang:2002yy}) estimated. 


\noindent
{\bf Acknowledgments}\\
The author would like to acknowledge very pleasant 
collaborations with B.A. Kniehl, A.A. Penin, V.A. Smirnov and M.  
Steinhauser on which parts of the work reported here are based.

\end{document}